\documentclass[twocolumn,aps,preprintnumbers,superscriptaddress]{revtex4}
\usepackage{graphicx}
\usepackage{bm}
\usepackage{amsmath, amsthm, amssymb}

\begin{document}

\title{Exploring matter wave scattering  by means of the phase diagram}

\author{Jeng Yi Lee}
\affiliation{
Institute of Photonics Technologies, National Tsing-Hua University, Hsinchu 300, Taiwan}
\affiliation{Department of Physics, National Chung-Hsing University, Taichung 402, Taiwan}

\author{Ray-Kuang Lee}
\affiliation{
Institute of Photonics Technologies, National Tsing-Hua University, Hsinchu 300, Taiwan}
\affiliation{Physics Division, National Center of Theoretical Science, Hsinchu 300, Taiwan}
\date{\today}

\begin{abstract}
For matter wave scattering from passive quantum obstacles,  we propose a phase diagram in terms of phase and modulus of scattering coefficients to explore all possible directional scattering patterns.
In the phase diagram, we can  not only have the physical bounds on  scattering coefficients for all channels, but also indicate the competitions among absorption, extinction, and scattering cross sessions.
With help of this phase diagram, we discuss different scenarios to  steer  scattering probability distribution, through the interference between $s$- and $p$-channels.
In particular, we reveal the required conditions to implement a quantum scatterer, i.e., a quantum dot in semiconductor matrix, with a minimum (or zero)  value in the scattering probability toward any  direction.
Our results provide a guideline in designing quantum scatterers with controlling and sensing matter waves.
\end{abstract}

\maketitle


\section{Introduction}
A deep understanding on the  wave scattering in  passive obstacles plays a crucial role in nanophotonics, with a variety of practical applications from nanoantenna, metasurface, sensor, imaging system,  to light harvesting ~\cite{science1,nature1}.
Exotic scattering phenomena at the subwavelength scale have been revealed such as coherent perfect absorbers, superscatterers, invisible cloaks, directional radiation scatterers, and superabsorbers~\cite{absorber,superscattering,cloak,kerker,forwardexp,backwardexp,superabsorber}.
To have a universal picture on all the  allowable scattering coefficients, we apply  the concept of energy conservation  law to absorption cross section, resulting in a phase diagram for electromagnetic waves~\cite{phase}.
Regardless of details on the geometric configurations and material properties, physical boundary and limitation for zero backward/forward scatterings can be easily illustrated in the phase diagram~\cite{kerker1}.

Similar to the classical counterpart, discussions on quantum particles encountering collisions have been an extensive research subject with different  physical disciplines.
For ultracold atoms of $^{87}$Rb, the Fano-signature interference in quantum scattering, between resonant $d$-wave  to the background $s$-wave, was observed experimentally~\cite{image1,image2}.
Through the classical-quantum correspondences between optical and matter waves,
the concept for invisible claoking has been applied to quantum scatterers formed by spherical quantum dots in core-shell heterostructures~\cite{qcloak1, qcloak2, thermoelectric, qcloak4}.
Moreover, it was realized that even though the electron dynamics in graphene is governed by the relativistic massless Dirac equation, the approach to deal with the related scattering problem of electronic transport in graphen   is similar to that used in  electromagnetic Mie theory~\cite{qcloak3, graphene1,graphene2}.
Under the scattering picture, Klein tunneling, with a complete suppression in the backward direction, can be transformed as an isotropic scattering process with spin-orbital interactions~\cite{graphene3}.

To design quantum resonant scatterers at subwavelength scale,  detailed physical quantities, such as effective mass and potential of a quantum particle, have been taken into account through higher-oder expansions~\cite{resonant}.
Nevertheless,  a systematic way to have all possible scattering states from quantum matter waves, irrespective of  configuration in a scattering system, is still lack.
In this work, we consider the inelastic scattering from quantum matter waves, based on   Schr\"{o}dinger equation.
In terms of the phase and modulus of scattering coefficients for passive quantum obstacles,
we propose a phase digram to provide complete information among extinction, absorption and scattering cross sections.
With probability conservation, i.e., the unitary relation, we discuss  directional scatterings of matter waves  by means of the phase diagram.
Unlike the electromagnetic counterpart ~\cite{kerker}, a perfect zero forward scattering can be achieved in quantum particles through  the interference between $s$- and $p$-waves when they have the same phases and modulus.
Moreover, a systematic way to realize scattering matter wave  with a reduced scattering probability at any arbitrary direction is also demonstrated.
Our results provide the guideline to design directional quantum scatterer with matter waves.

\begin{figure}[t]
\centering
\includegraphics[width=8.4cm]{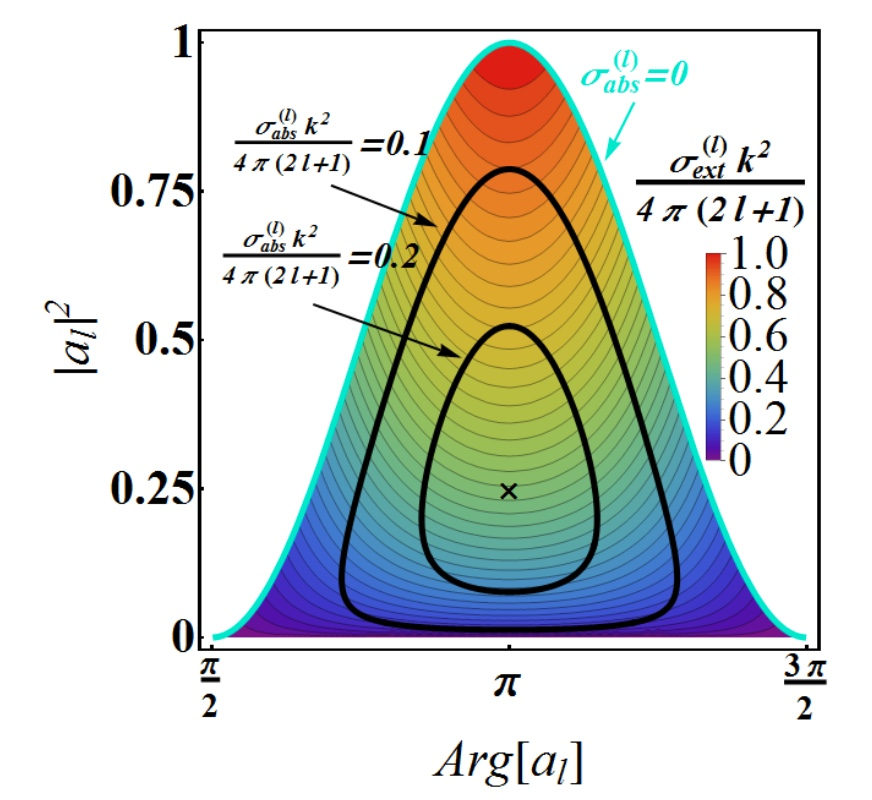}
\caption{Phase diagram for quantum matter waves, defined by the phase $Arg[a_l]$ and modulus $\vert a_l \vert^2$ of scattering coefficient $a_l$, in each angular momentum channel labeled by the index $l$.
Here, the contour-plot shows the value of normalized extinction cross section,  defined as $\sigma_{ext}^{(l)}k^{2}/4\pi(2l+1)$ with the range $[0, 1]$.
Two different sets of  constant normalized absorption cross section, defined as $\sigma_{abs}^{(l)}k^{2}/4\pi(2l+1)$  bounded within $[0, 0.25]$, are depicted in black-curves.
The maximum value in the normalized absorption cross section is also marked by the black-cross, i.e. $\sigma_{abs}^{(l)}=0.25$ located at $(\text{Arg}[a_{l}], \vert a_{l} \vert^{2})=(\pi, 0.25)$. }
\end{figure}
 
\section{Phase diagram for quantum scattering matter waves}
For a quantum particle with energy $E > 0$ and momentum along the $z$-direction,  in  absence of a finite-ranged obstacle, the unbounded stationary eigen-states can be described by a plane wave, $\psi_{i}=e^{ikz}$, with the  corresponding wavevector $\vec{k}=k\hat{z}$.
Even though the introduction of a quantum obstacle may alter the original eigen-states, the corresponding  eigen-energy  remains unchanged  in the elastic process.
Then, one can apply the Lippmann-Schwinger equation to construct the new scattered state for  a finite-ranged obstacle in an arbitrary shape~\cite{jj}.

Furthermore, if the scattering obstacle possess rotational invariance, the Hamiltonian for the quantum matter wave commutes with angular momentum operators, i.e., $\vec{L}^{2}$ and $L_{z}$, corresponding to  the total and $z$-component angular momentum operators, respectively.
With this symmetry, we can apply the partial wave formalism  to matter wave scattering problem.
In the asymptotic region, one can write down the wave function for quantum obstacle as~\cite{book1} 
\begin{equation}\label{asymptotic}
\psi(\vec{z})=e^{i\vec{k}\cdot\vec{z}}+\frac{e^{ikr}}{r}f(\theta).
\end{equation}
Here, we have  incident matter wave as a plane wave, and the corresponding scattering amplitude  $f(\theta)$ can be defined as
\begin{equation}
f(\theta)=-\frac{i}{k}\sum_{l=0}^{l=\infty}(2l+1)a_{l}P_{l}(\cos\theta),
\end{equation}
with the scattering coefficient $a_{l}$, the Legendre polynomial $P_{l}$, and the index in angular momentum channels labeled as $l$.
It is noted that a monopole ($l=0$) is also called as $s$-wave; while a dipole ($l=1$) is called as $p$-wave.
From the asymptotic result shown in Eq.(\ref{asymptotic}),
by integrating probability flux along the radial component $r$ over a closed area, we can directly find the corresponding absorption (also called as the reaction), scattering, and extinction cross sections, respectively,
\begin{eqnarray}
\sigma_{abs}&=&-\frac{4\pi}{k^{2}}\sum_{l=0}^{l=\infty}(2l+1)\{\vert a_{l}\vert^{2}+Re[a_{l}]\}\\
\sigma_{scat}&=&\frac{4\pi}{k^{2}}\sum_{l=0}^{l=\infty} (2l+1)\vert a_{l}\vert^{2}\\
\sigma_{ext}&=&\sigma_{abs}+\sigma_{scat}.
\end{eqnarray}
One can see that in  these convergent series, the dominant terms  are often determined by  environment size parameters, i.e., $k a$ with the effective range of scatterer radius $a$ ~\cite{qcloak1,qcloak2,qcloak3}. 
We want to remark that through the optical theorem, the  extinction cross section is also  linked to  the scattering amplitude in the forward direction~\cite{jj} $\sigma_{ext}=4 \pi \, \text{Im}[f(0)]/k$.
In addition, an alternative method to calculate the cross section is using the phase shift $\delta_{l}$, by re-defining the scattering coefficient as $2a_{l}+1=\text{exp}[i\delta_{l}]$ ~\cite{book1}.

With probability conservation, i.e., the unitary relation, one can deduce that $\sigma_{abs}^{(l)}=0$ (or $\sigma_{scat}^{(l)}=\sigma_{ext}^{(l)}$) for each angular momentum channel.
Here,  $\sigma_{abs}^{(l)}$ is the partial absorption cross section for the $l$-th channel, defined as $-4\pi(2l+1)\{\vert a_{l}\vert^{2}+Re[a_{l}]\}/k^{2}$.
Then, one can see that the unitary relation  guarantees a real value for phase shift,  $\delta_{l}$.
On the contrary,  if some incident quantum particles are annihilated by scattering obstacles or lose their energy, we have $\sigma_{abs}^{(l)}> 0$ for a non-zero absorption cross section.
Next, we introduce a phasor representation for the scattering coefficients, by writing $a_{l}=\vert a_{l}\vert\, \text{exp}\{i\, Arg[a_{l}]\}$, into this inequality $\sigma_{abs}^{(l)}> 0$.
As a result, a phase diagram for every partial extinction cross section emerges naturally, as shown in Fig. 1, defined by the phase $Arg[a_l]$ and modulus $\vert a_l \vert^2$ of scattering coefficient $a_l$, in each angular momentum channel labeled by the index $l$.
Here, we note that in this phase diagram, the vertical axis gives the strength in the scattering channel; while the horizontal axis reflects the  phase in  scattering coefficient.
Moreover, colored region reveals all the allowable solutions for a passive  quantum obstacle, i.e., $\sigma_{abs}^{(l)}\geq 0$; while the uncolored (or in white-color) regions correspond to unallowable solutions.
The range for allowable solutions  is  $[\pi/2, 3\pi/2]$ for phase, and $[0, 1]$ for modulus.
Moreover, between the colored  and uncolored regions, the boundary is depicted by  a trajectory in light-blue-color, which exactly follows the unitary relation. 

In addition to the allowable solutions for scattering coefficients,  in a single phase diagram, one can also display information among the normalized absorption, scattering , and  extinction cross sections, defined as $\sigma_{abs}^{(l)}k^{2}/4\pi(2l+1)$, $\sigma_{scat}^{(l)}k^{2}/4\pi(2l+1)$, and $\sigma_{ext}^{(l)}k^{2}/4\pi(2l+1)$, respectively.
As shown in Fig. 1, the value   of normalized extinction cross section is depicted in the contour-plot, which is bounded within $[0, 1]$.
Then, we also give two sets of constant absorption cross section in black-colors, $\sigma_{abs}^{(l)}k^{2}/4\pi(2l+1)= 0.1$ and $0.2$, respectively.
These black-curves reveal non-trivial scattering events, with a constant absorption cross section, but totally different extinction and scattering cross sessions.
The maximum value in the absorption cross section is $0.25$, see the black-cross marker, which reflects the resonance condition for a complete cancellation in the out-going spherical matter wave, as $a_{l}=-1$ or $(\text{Arg}[a_{l}], \vert a_{l} \vert^{2})=(\pi, 0.25)$. 
Exotic scattering phenomena, such as  quantum sensors~\cite{sensor} and anti-laser~\cite{antilaser} can also be clearly illustrated in this phase diagram~\cite{phase}.
In addition,  the common adopted Born approximation to deal with weak scattering, valid for a shallow potential or a small particle size compared to the wavelength in quantum wave, the  phase in the scattering coefficient is always $\pi/2$ or $3\pi/2$ ~\cite{comment1}.
Moreover, even though one can introduce a complex potential into Schr\"odinger equation or impose an imaginary term for the phase shift ~\cite{book1, book2}, such a phenomenological approach is  beyond the scope of this work.

 \section{Interferences from dominant  $s$- and $p$-waves}

\begin{figure*}[t]
\begin{center}
\includegraphics[width=12.0cm]{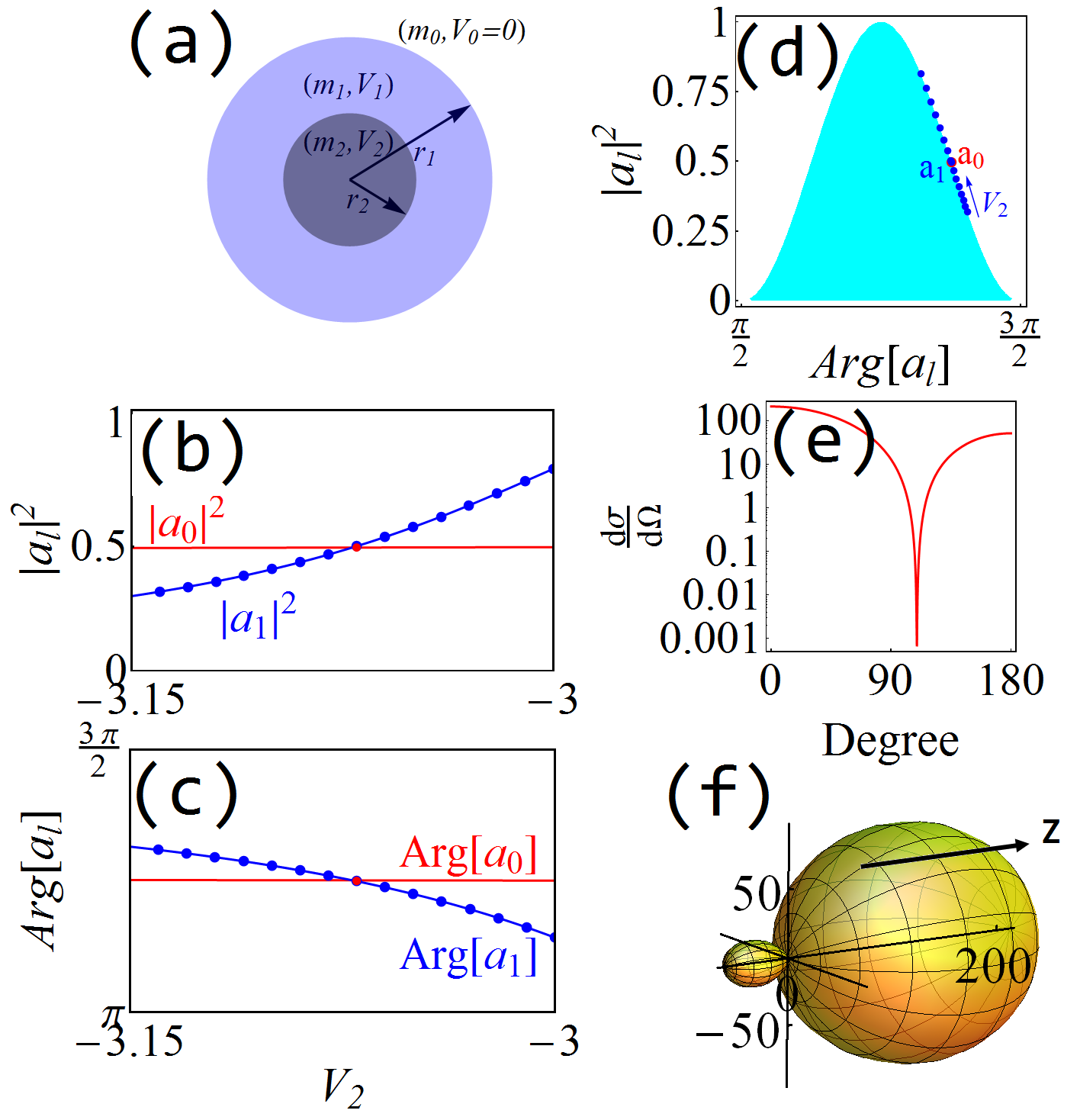}
\end{center}
\caption{(a) Illustration of our quantum scatterer, formed by a two-layered quantum dot in the shape of a core-shell sphere. 
The corresponding (b) modulus and (c) phase  for the the scattering coefficients: $a_0$($s$-wave) and $a_1$ ($p$-wave), in red- and  blue-colors, respectively, as a function of the effective potential in the core region, $V_{2}$.
Note that both in (b) and (c), two curves intersect at the same point with  $V_{2}=-3.07$ eV.
(d) The trajectories in phase diagram for the solutions shown in (b) and (c). 
The differential scattering cross section as a function of angles are shown in (e) log-plot and (f) 3D-plot, with a clear dip (node) at  $\theta=109.5^{\circ}$.
The other parameters used are $r_{1}=4$ nm, $r_{2}=0.4 r_{1}$,  $E=0.022$ eV, $V_{1}=-2.83$ eV, $m_{0}=0.065$ $m_{e}$, $m_{1}=0.034$ $m_{e}$, and $m_{2}=0.021$ $m_{e}$, respectively.}
\end{figure*}
 
Although in the first look, this phase diagram for quantum matter waves, shown in Fig. 1, shares the same similarity to that introduced for electromagnetic system, the underline physical interpretations are totally different.
In quantum matter wave formalism, the physical measurement for  probability is implemented through an ensemble average.
Moreover, instead of electric or magnetic dipole excited in electromagnetic waves by subwavelength structures, the two lowest-orders for predominant scattering phenomena in quantum subwavelength system are monopole and dipole, i.e., $s$-wave with $l = 0$ and $p$-wave with $l = 1$, respectively.

Now, at the subwavelength scale, we study the differential scattering cross section $d\sigma_{scat}/{d\Omega}$ for a quantum scatterer by calculating
\begin{equation}\label{differential}
\frac{d\sigma_{scat}}{d\Omega}=\vert f(\theta)\vert^{2},
\end{equation}
which means the ensemble average in the probability scattered into the angle $\theta$ ~\cite{jj}.

In particular, we seek for directional scattering events by asking for the zero probability distribution through the superposition from  dominant  $s$- and $p$-waves.
That corresponds to find a family of solutions for the scattering coefficients $a_0$ and $a_1$ to satisfy the following condition:
 \begin{equation}
 a_{0}+3\, a_{1}\, \cos\theta = 0,
 \end{equation}
 or equivalently in the phasor representation:
\begin{equation}\label{sp}
\vert a_{0}\vert e^{i\text{Arg}[a_{0}]}+3\vert a_{1}\vert e^{i\text{Arg}[a_{1}]}\vert \cos\theta\vert e^{i\text{Arg}[\cos\theta]}= 0.
\end{equation}
Here, $\theta$ is defined in the spherical coordinate, which is bounded by $[0,\pi]$.
In the regime $\theta = [0, \pi/2]$, the argument of $\cos\theta$ would be $0$; while  when  $\theta = [\pi/2,\pi]$, the argument is $\pi$.

Instead of trivial solutions, $a_0 = a_1 = 0$, let us consider the scenario when two scattering coefficients are the same both in their phase and modulus.
 Then, to satisfy the required condition given in Eq. (7), we have 
$1+3\cos\theta=0$, or equivalently $\theta =109.5^{\circ}$, which reveals a node (zero) in the corresponding probability distribution.
It should be remarked that in this scenario, once the phase and modulus for $s$ and $p$ wave are the same, the system can possess any scattering cross section, as  indicated in Fig. 1.

For possible experimental implementation, we consider conducting electrons within a semiconductor matrix encountering an artificial quantum dot as an example.
This artificial scatterer is constituted by two concentric spheres,  with isotropic,  homogeneous effective masses and potentials in the shell and core regions, denoted as ($m_1$, $V_1$) and ($m_2$, $V_2$) as illustrated in  Fig. 2 (a), respectively.
Effects of edge roughness and Coulomb screening are neglected for the illustration.
With advance in fabrication technologies,  quantum dots with a controllable size and tunable energy-band structure demonstrated recently can be a good candidate as our quantum scatterer~\cite{quantumdot1}.
Here, we choose the system parameters from a single quantum dot as $r_{1}=4$ nm for the radius of  whole sphere, $r_{2}=0.4 r_{1}$ for the core region, $E=0.022$ eV for the incident energy,  and $V_{1}=-2.83$ eV for the effective potential in the shell region.
As for the effective mass outside the quantum scatterer, inside the shell region, and inside the core region, we set $m_{0}=0.065$ $m_{e}$, $m_{1}=0.034$ $m_{e}$, and $m_{2}=0.021$ $m_{e}$ (in the unit of electron mass $m_e$),  respectively.
Then, by scanning the effective potential in the core region, $V_2$, we report the
  changes in the modulus and phase for the corresponding scattering coefficients in the two lowest-orders, $a_0$ and $a_1$, as shown in the red- and blue-colored curves of  Fig. 2(b) and (c), respectively.
As one can see, in the range  $V_2 = [-3.15$ eV, $-3$ eV$]$, there exists an crossing point from two scattering coefficients, which is located at the same effective potential in the core region, i.e.,  $V_{2}=-3.07$ eV.
At this effective potential, we have exactly $a_{0}=a_{1}$, which gives the operation point to satisfy a directional scattering with a node at  $\theta=109.5^{\circ}$.

For a lossless system, our Hamiltonian ensure a unitary transform, guaranteeing the scattering coefficients $a_{0}$ and $a_{1}$ localed in the light-blue-colored trajectory shown in Fig. 1.
Regardless of material parameters and geometry size,  we have $\sigma^{(l)}_{abs}=0$ due to the unitary relation.
To illustrate such a trajectory along the boundary between allowable and unallowable regions, in Fig. 2(d), we depict the locations from the solutions shown in Figs. 2 (b-c) in the phase diagram.
As one can see that, in this chosen range,  $V_2 = [-3.15$ eV, $-3$ eV$]$, the corresponding modulus of  scattering parameter for $p$-wave, $\vert a_1\vert$, increases as $V_2$ increases; while the modulus for $s$-wave, $a_0$, remains almost unchanged ~\cite{note}.

The advantage of phase diagram includes not only the integrated information of phase and modulus for each scattering channel,  but also the corresponding  cross sessions at the same map.
For the crossing point at $V_{2}=-3.07$ eV, we also calculate the  corresponding differential scattering cross section as a function of angles, as shown in Fig. 2(e) and (f) for the log-plot and 3D-plot, respectively.

\begin{figure*}[t]
\begin{center}
\includegraphics[width=16.0cm]{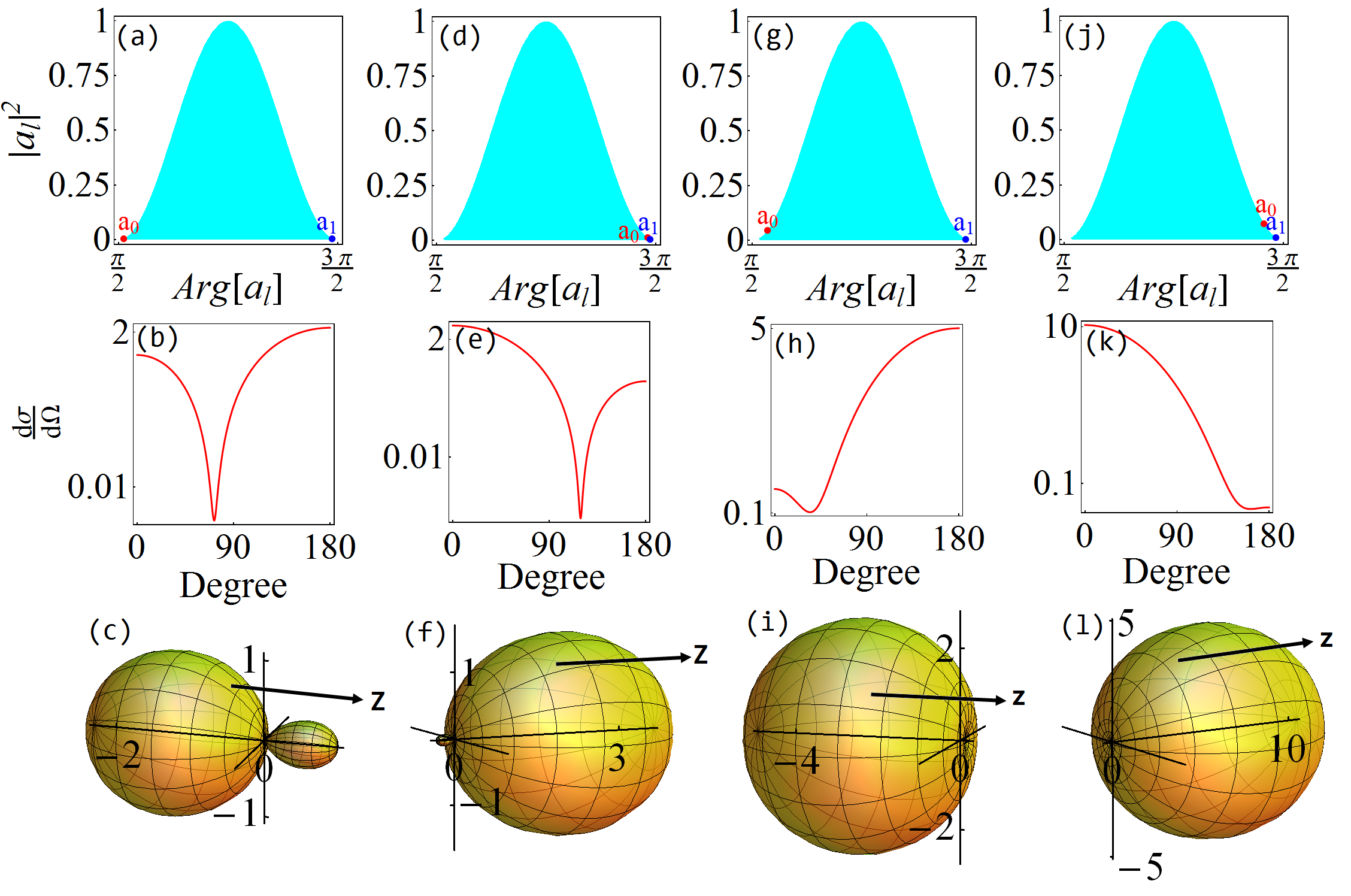}
\end{center}
\caption{Four selective-angles to minimize directional scattering cross section. The corresponding locations of $s$ and $p$ in the phase diagram are shown in the first row (a, d, g, j), corresponding a minimum (or zero) value in the scatting angle  at $\theta=72^{\circ}$, $120^{\circ}$, $0^{\circ}$, and $180^{\circ}$. The resulting differential scattering cross sections as a function of angle are depicted in the second and third rows, for the log-plots and 3D plots, respectively.}
\end{figure*}

Now, we go one step further  by considering the case $a_{0}\neq a_{1}$ in two scattering channels.
For $\theta\in [0,\pi/2]$, the argument of $\cos\theta$ is $0$.
Thus, in order to minimize Eq.(\ref{sp}), the arguments between $a_{0}$ and $a_{1}$ need to be out-of-phase, in orer to form a destructive  interference.
As clearly illustrated in the phase diagram, such a out-of-phase condition can only be satisfied when this scattering obstacles is asked to support scattering events located in the two opposite boundaries.
The corresponding condition for modulus to satisfy  Eq. (\ref{sp}) becomes
\begin{equation}\label{magnitude_sp}
\begin{split}
\frac{\vert a_{0}\vert}{\vert a_{1}\vert}&=3\vert\cos\theta\vert.
 \end{split}
\end{equation}
As an example, in the first column of Fig. 3, we demonstrate the scattering pattern with a minimization in the scattering distribution at  $\theta=72^{\circ}$.
The corresponding locations of $s$- and $p$-wave channels  in the phase diagram  are  depicted in Fig. 3(a); while  the resulting scattering distributions are represented in log-plot and 3D-plot, i.e.,  Figs. 3(b) and (c), respectively.
Since the phase difference between $a_{0}$ and $a_{1}$ is almost $\pi$, one can expect to have a node in the scattering cross section for $\theta\in [\pi/2,\pi]$.
In this example, the system parameters used are all the same as those shown in Fig. 2(f), but with $V_{1}=1.2$ eV and $V_{2}=-2.48$ eV.

For $\theta\in [\pi/2,\pi]$, the argument of $\cos\theta$ is $\pi$.
As a result, Eq.(\ref{sp}) can be reduced to the condition $\vert a_{0}\vert e^{i\text{Arg}[a_{0}]} = 3\vert a_{1}\vert e^{i\text{Arg}[a_{1}]}\vert \cos\theta\vert$. 
Then, only when $\theta\in [\pi/2,\pi]$, we can have a minimum (or zero) value in the corresponding scattering distribution.
Moreover, from the phase diagram shown in Fig. 3(d), we can see that both $a_{0}$ and $a_{1}$ locate at the same side, minimizing their phase difference.
Due to the same sign in the phases of $a_{0}$ and $a_{1}$, the resulting scattering pattern is enhanced along the forward direction.
In Figs 3. (e-f), we report the differential scattering cross section  with a dip (node) at $\theta=120^{\circ}$.
Here,  the parameters are all the same as those in Fig. 2 (f), but with $V_{1}=1.2eV$ and $V_{2}=-2.49eV$.

It is known that for electromagnetic waves, Kerker {\it et al.} proposed to generate zero backward scattering (ZBS) and zero forward scattering (ZFS) through a proper combination of magnetic and electric dipoles~\cite{kerker}.
However, due to the optical theorem, there is  always a residue in the forward scattering direction for any passive electromagnetic scatterer.
Similar to the optical theorem,   in the forward direction $\theta = 0$, the corresponding scattering amplitude $f(\theta = 0)$, shown in  Eq. (6),  is also  linked to the extinction cross section.
Instead of a perfect ZFS, we can only minimize the scatting distribution at the  angle $\theta = 0$.
To do this, by fixing $\theta = 0$, one can perform the minimization for $f(0)$ with the conditions $\vert a_{0}\vert=3\vert a_{1}\vert$ and $\text{Arg}[a_{0}]-\text{Arg}[a_{1}] = \pm \pi$.
As shown in the third column in Fig. 3 (g-i), we report the scattering pattern for matter waves with a nearly zero value in the  forward scatting by changing the  parameters $V_{1}=1.2$ eV and $V_{2}=-2.492$ eV.
Furthermore,  ZBS can be easily generated with a family of scattering events in the phase diagram~\cite{kerker1}.
A typical example for ZBS is illustrated in  Fig. 3(j-k) with $V_{1}=-1.8$ eV and $V_{2}=-0.8$ eV.

Before conclusion, we want to stress that although our discussion is limited to plane wave formalism of quantum scatterers, the partial wave decomposition is independent from the incident wave form.
As for other geometry in the quantum scatterer, one may also find a suitable orthogonal basis to decompose the incident and scattered waves.
All the physical principles we apply here is to embed the law of  passivity on absorption cross section.

 \section{Conclusion}
 In conclusion, by considering inelastic scattering for quantum matter waves with the condition $\sigma^{(l)}_{abs}>0$, we propose a phase diagram for every angular momentum channel.
 In a single phase diagram, we can reveal not only the physical bounds on phase and modulus of scattering coefficients, but also indicate the  competitions among absorption, extinction and scattering cross sessions.
With the help of this phase diagram, we discuss different scenarios through the interference between  $s$-  and $p$-waves, with the demonstrations to steer the scattering probability distribution from quantum matter waves.
 We find that with the same phase and modulus from dominant channels, one can have a zero value in the scattering probability distribution at $\theta=109.5^{\circ}$.
 However, as the strength in the $s$- and $p$-waves are different, we can also minimize the scatting probability distribution at a give angle.
With an artificial core-shell quantum dot in semiconductor matrix as an example,  such a systematic way by means of the phase diagram is able to provide guidelines in designing quantum scatterers. 

 \section*{Acknowledgments}

This work has been supported by Ministry of Science and Technology,  Taiwan, under Contract MOST 105-2686-M-007-003-MY4.


\begin{thebibliography}{99}
\bibitem{comment1}
In the first-order Born approximation, the corresponding scattering amplitude for a general potential  is always a real number,  but its sign (negative or positive) depends on the  barrier or well in  potentials ~\cite{book1}, resulting in the phase of scattering coefficients being $\pi/2$ or $3\pi/2$.
\bibitem{note}
It should be noted that in this case,  by tuning $V_2$ $p$-wave can be excited to be on resonance to the background $s$-wave, forming a Fano-signature in the scattering cross section.
\end{thebibliography}
 \end{document}